\documentclass[10pt]{article}
\usepackage[utf8]{inputenc}
\usepackage{setspace}
\usepackage[margin=1in]{geometry}
\usepackage{graphicx}
\graphicspath{ {./figures/} }
\usepackage{subcaption}
\usepackage{amsmath}
\usepackage{lineno}
\usepackage{hyperref}
\usepackage{helvet}
\usepackage[table,xcdraw]{xcolor}
\usepackage{wrapfig}
\usepackage{placeins}

\usepackage[style=nejm, 
citestyle=numeric-comp,
sorting=none]{biblatex}
\title{\textbf{Precise Magnetic Field Mapping of the EMPHATIC Phase 1 Magnet with COMSOL}}

\date{}

\begin{document}

\maketitle
\vspace{-5em}
\begin{center}
    
Prachi Sharma$^{*1}$ for the EMPHATIC Collaboration$^{\dag}$

$^1$\textit{Department of Physics, Panjab University, Chandigarh, IN}\\
$^{\dag}$\textit{Fermilab, Batavia, IL, US}

\end{center}

\begin{abstract}
\begin{center}
A compact Halbach array magnet is used to measure the momentum of the secondary particles in \textbf{EMPHATIC} (\textbf{E}xperiment to \textbf{M}easure the \textbf{P}roduction of \textbf{H}adrons \textbf{A}t a \textbf{T}est beam \textbf{I}n \textbf{C}hicagoland)
. Hall probe data was taken for the central cylindrical bore of the magnet and a field map was constructed. COMSOL Multiphysics® Software is used for modeling the magnet and constructing the corresponding magnetic field map. We present a fitting approach where the hall probe data is used to determine a 1mm-spacing map of the entire volume of the magnet using COMSOL. The new map will allow for linear interpolation within the volume, and expand the map to outside the measurement volume, thus increasing the acceptance and precision of EMPHATIC’s tracking system.\\

\end{center}
\end{abstract}

\begin{center}

\textit{Contribution to the 25th International Workshop on Neutrinos from Accelerators- NUFACT 2024,\\ Argonne National Laboratory, US.}

\end{center}

\section{Introduction}
EMPHATIC~\cite{Pavin_2022} is a Fermilab-based table-top size experiment focused on hadron production measurements. Flux is a limiting systematic for all neutrino cross-section measurements by current experiments and we rely on a-priori predictions of the flux for analyses, including measurements of neutrino oscillations, neutrino-nucleus cross sections, and beyond-the-Standard Model searches. These flux predictions depend on simulations of hadron production and focusing, which introduce 10-20\% uncertainties. The experiment aims to improve our understanding of hadron scattering and production with better than 10\% measurements, and provide the first-ever hadron spectrum measurement downstream of a target and horn. \\

The EMPHATIC Phase 1 magnet is a 3-layer cylindrical Halbach dipole made of 48 N52 Neodymium magnets. We present the process of generating a 1 mm resolution magnetic field map using COMSOL modeling, fitting it to measured data. This study aims to achieve higher precision than the previous AP-STD mapping with Hall probe measurements, addressing the need for better track separation, improved momentum and angle resolution, and enhanced detector acceptance.

\section{EMPHATIC Magnet and Magnetic Field Mapping}

\begin{figure}[h]
    \centering
    \begin{minipage}{0.6\textwidth}
        \centering
        \includegraphics[width=0.4\linewidth]{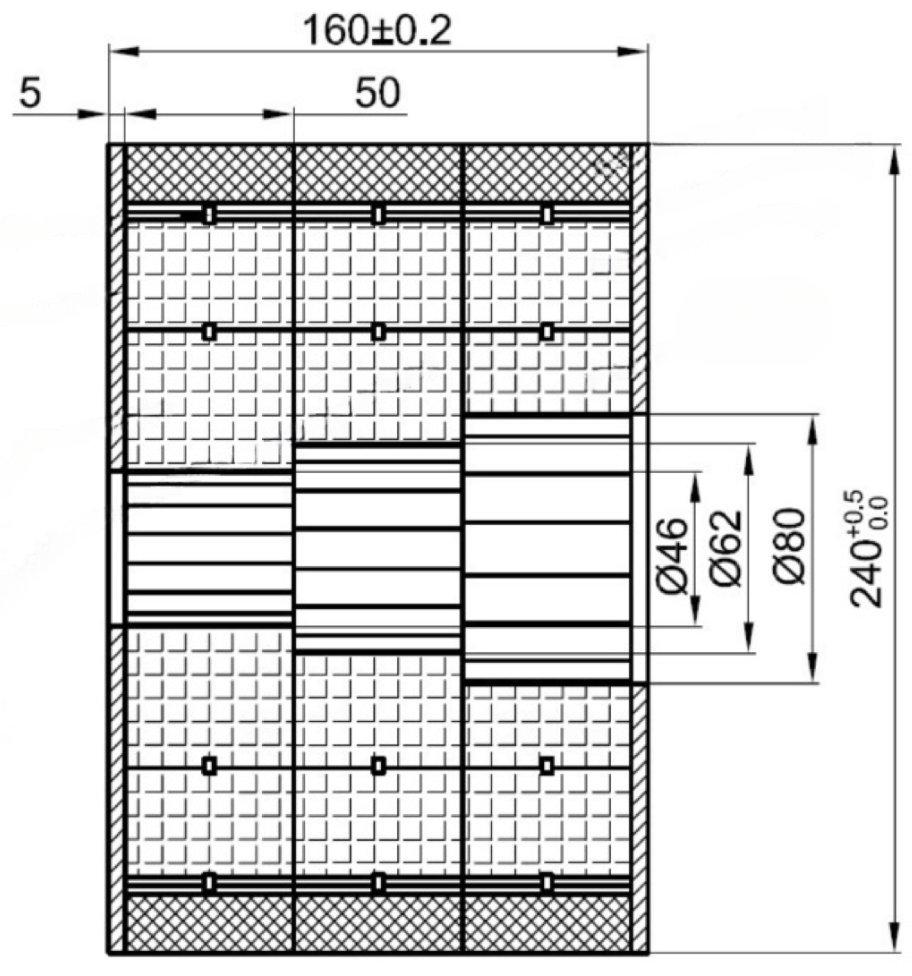}
        \includegraphics[width=0.4\linewidth]{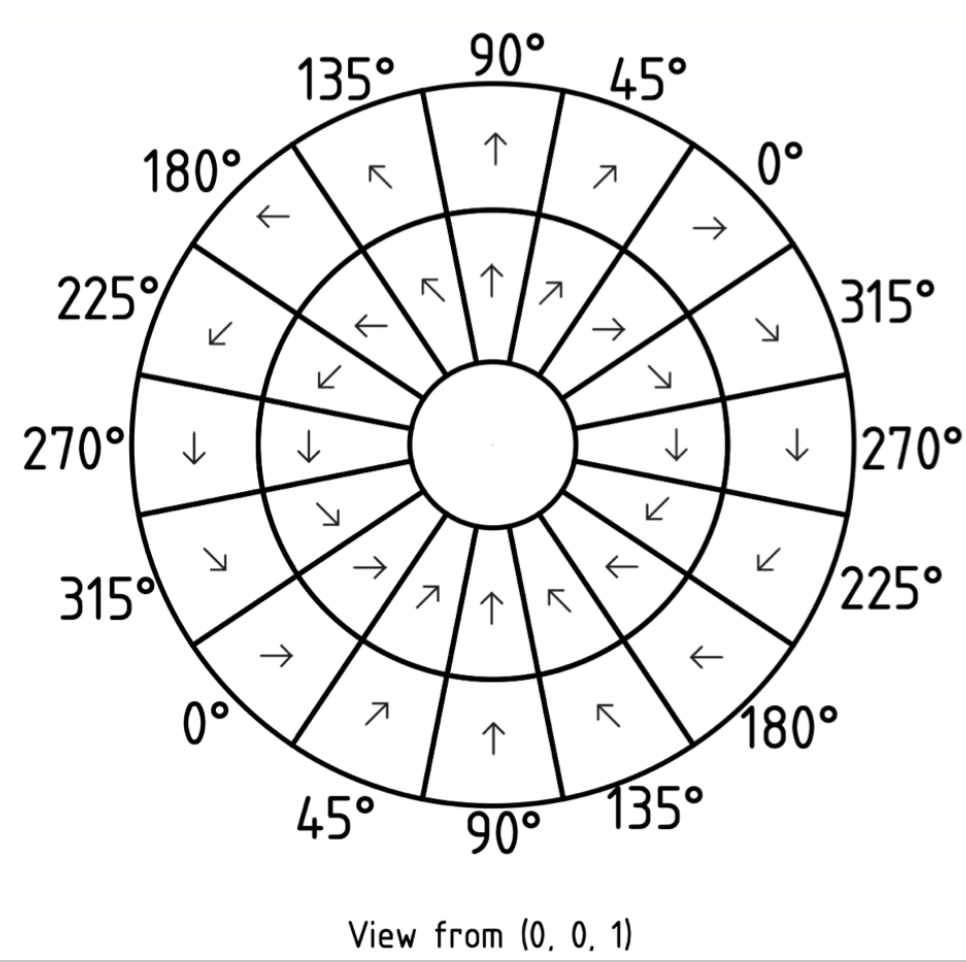}
    \end{minipage}\hfill
    \begin{minipage}{0.4\textwidth}
        \centering
        \caption{\textbf{(Left)} Schematic of the 3-layer EMPHATIC Phase 1 magnet.\\
        \textbf{(Right)} Magnetisation directions of the 16 NdFeB components in each layer.}
        \label{fig_halbach_array}
    \end{minipage}
\end{figure}

\begin{figure}[h]
    \centering
    \includegraphics[width=0.9\textwidth]{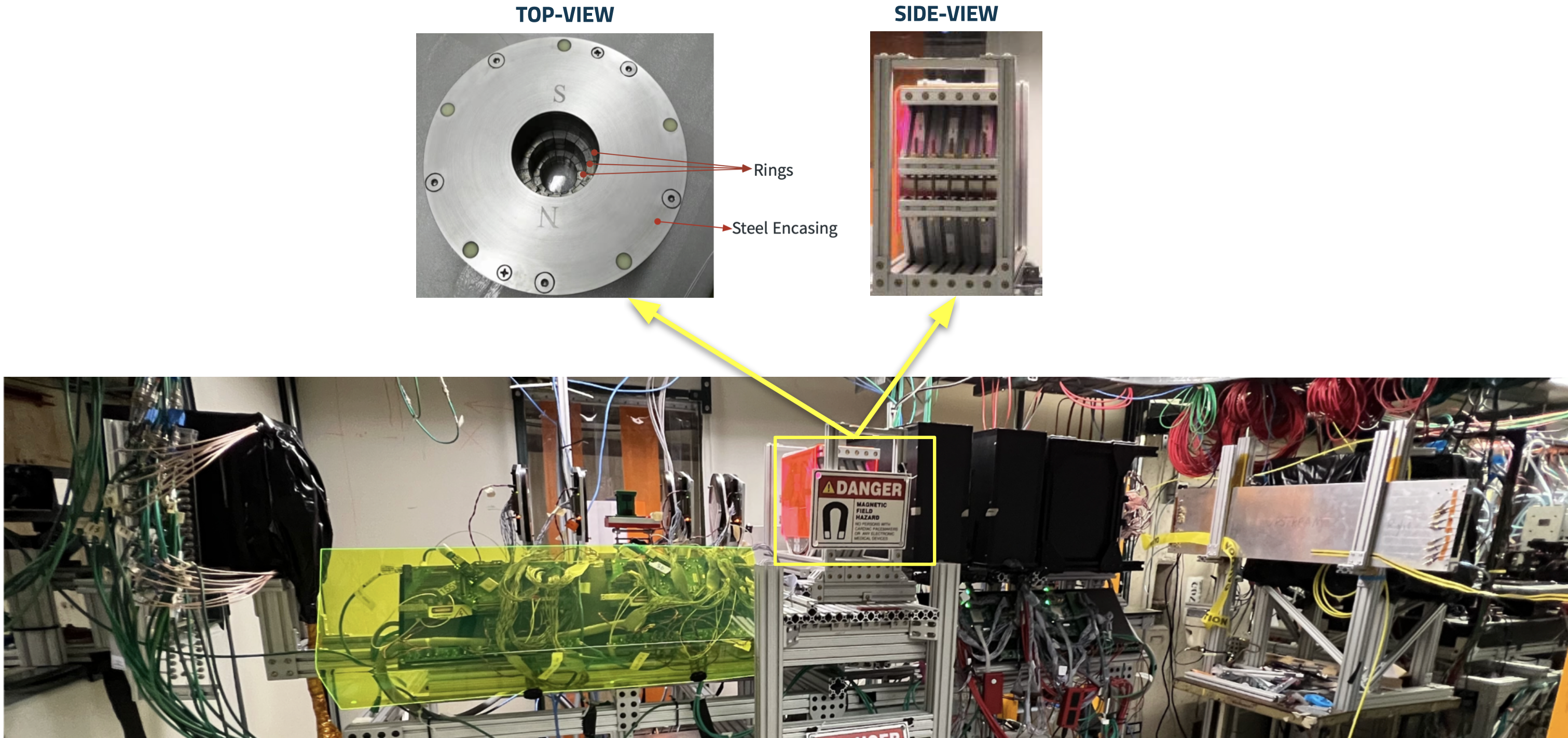}
    \caption{Images of the EMPHATIC Phase 1 magnet from the top and side views within the EMPHATIC Phase 1 run setup.}
    \label{fig:emphatic_magnet}
\end{figure}

The EMPHATIC Phase 1 magnet consists of three cylindrical Halbach layers, each 50 mm thick with 5 mm steel cladding. The layer diameters are 46 mm, 62 mm, and 80 mm, as shown in Figure \ref{fig_halbach_array}. Each of the NdFeB segments has a magnetic field strength of 1.44 T, with an integrated field of 0.2 Tm~\cite{bellantoni2024}. The magnet spans coordinates (0,0,0) mm to (0,0,160) mm (Figure \ref{fig:emphatic_magnet}).\\

Magnetic field measurements, conducted by Fermilab’s AP-STD in March 2023, used a SENIS 3-axis magnetic transducer (sensitivity $5V/T$)~\cite{emphatic1580} with a Hall probe offering 150 × 150 $\mu m^2$ spatial resolution. Data points were recorded at 5 mm intervals across a grid from -15 mm to +15 mm in the x and y directions and from -140 mm to 310 mm in the z direction.\\

\begin{figure}[h!]
    \centering
    \includegraphics[width=0.85\textwidth]{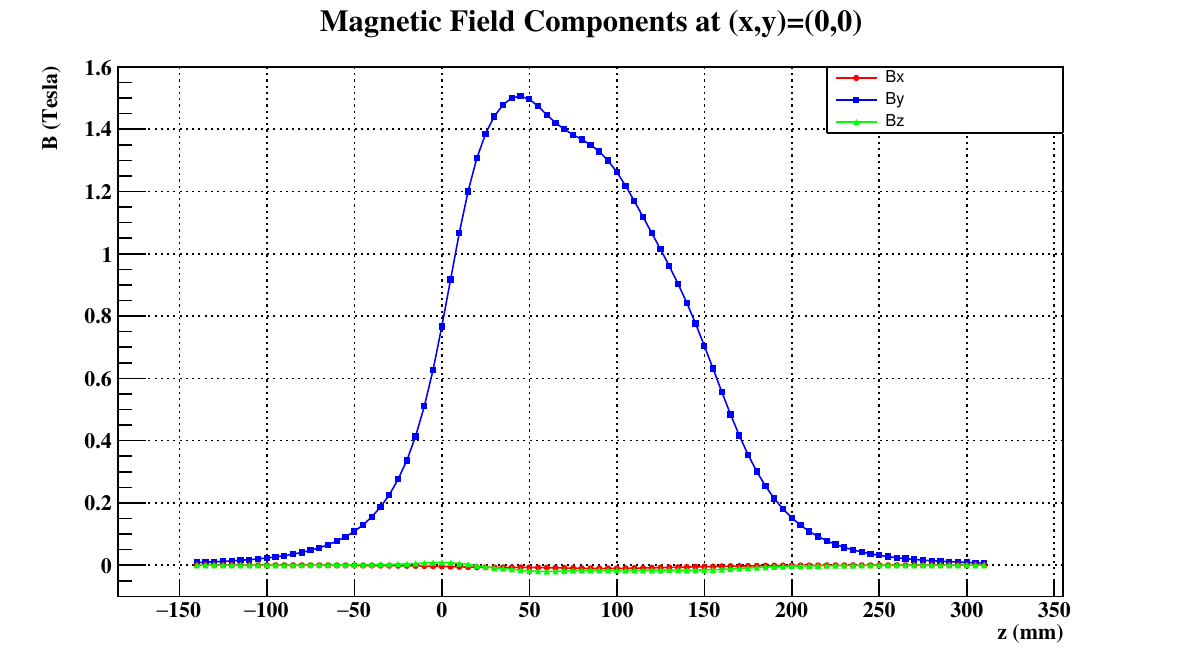}
    \caption{Magnetic field components (Bx, By, Bz) along the z-axis at (x, y) = (0, 0).}
    \label{field_plots}
\end{figure}

Figure \ref{field_plots} shows the magnetic field components (Bx, By, Bz) along the central z-axis (x=0, y=0), where the peak field of 1.506 T occurs at z=45 mm. Field maps at 45 scan locations reveal no sharp discontinuities but show minor asymmetries in theoretically symmetric regions.

\newpage

\section{Modeling the Magnet with COMSOL}
\begin{wrapfigure}{r}{0.3\textwidth}
    \centering
    \vspace{-30pt}
    \includegraphics[width=0.28\textwidth]{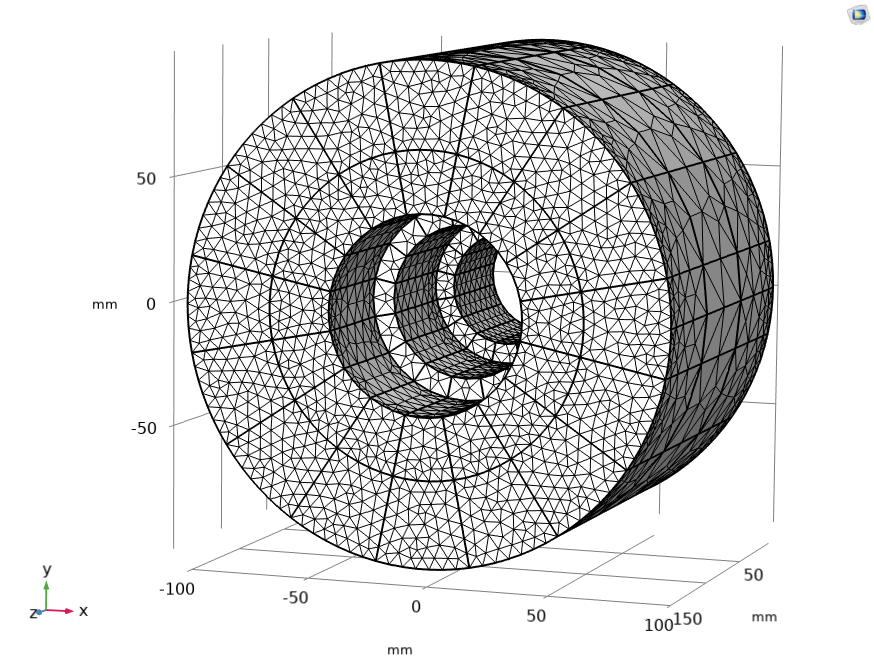}
    \caption{3D view of the magnet model in COMSOL.}
    \label{fig:comsol_model}
    \vspace{-10pt}
\end{wrapfigure}

COMSOL Multiphysics 6.1~\cite{comsol2024} was used to simulate the magnet system. The model employs an \textit{extra fine mesh} and the \textit{Magnetic Fields, No Currents (MFNC)} interface. The materials include \textit{Air}, \textit{NdFeB}, and \textit{Steel 304}, with initial magnetization set at \textbf{1.44 Tesla per component}. The magnet's center is located at \textit{(0, 0, 80) mm}, with offsets of \textit{\( X = 0 \, \mathrm{mm}, Y = 0 \, \mathrm{mm}, Z = 80 \, \mathrm{mm} \)}. A total of \textbf{147 fit parameters} were defined, comprising 144 magnetization parameters for the 48 components across three layers and 3 offset parameters (\(X\), \(Y\), \(Z\)).

\subsection{Data vs. Nominal Model}  
\begin{figure}[h!]
    \centering
    \includegraphics[width=0.8\linewidth]{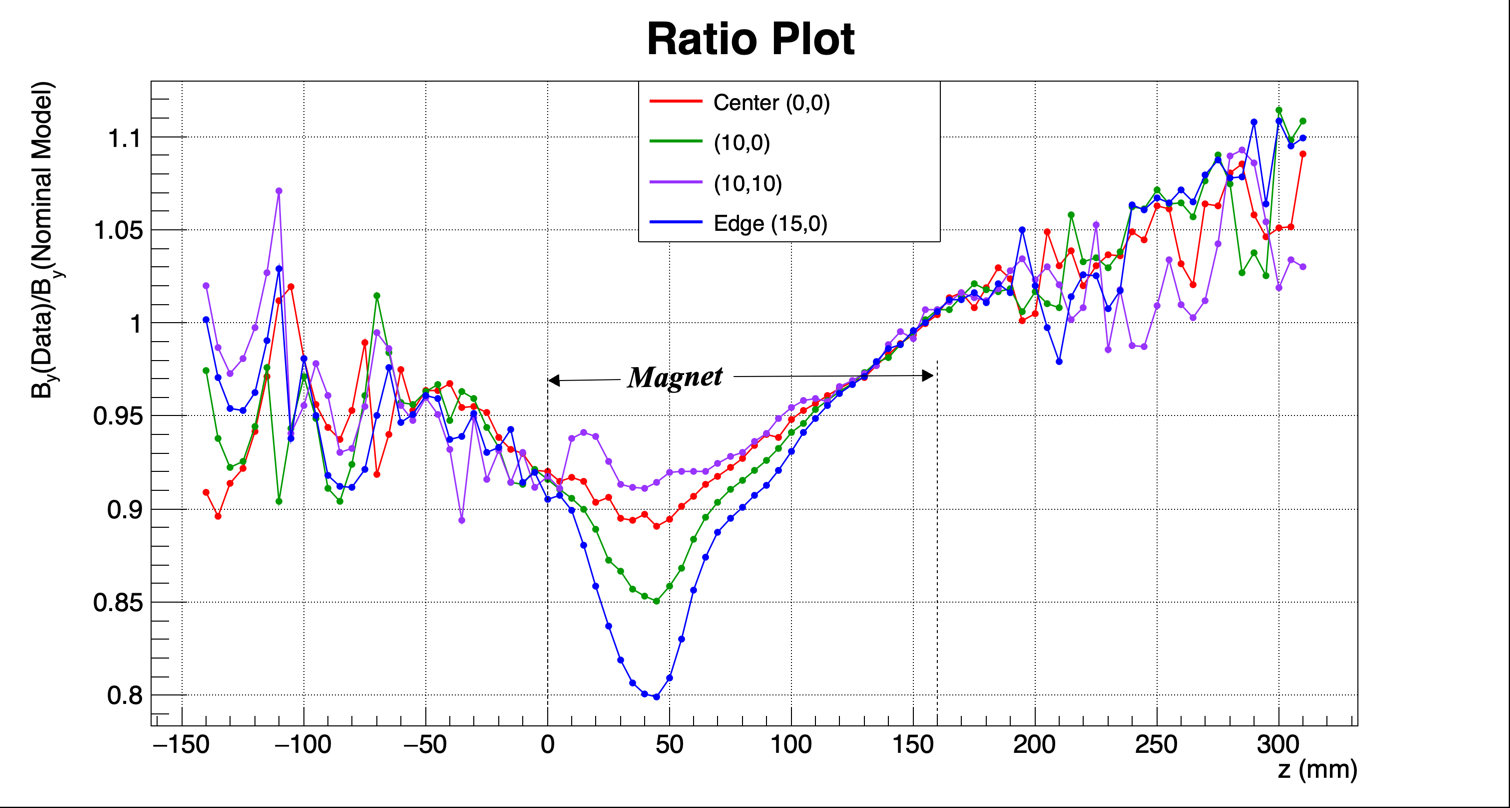}
    \caption{Ratio plot of data and COMSOL model at major positions, illustrating grid variations.}
    \label{fig_ratio_plot_data_nominal_model}
\end{figure}

To compare the data with the nominal COMSOL model, four positions—center \( (0, 0) \), \( (10, 0) \), \( (10, 10) \), and edge \( (15, 0) \)—were selected to represent variations across the grid. Figure~\ref{fig_ratio_plot_data_nominal_model} shows ratio plots at these positions, highlighting discrepancies between the data and the fit. At the \textcolor{red}{center (0,0)}, the maximum discrepancy is approximately \( 12\% \) at a \( z \)-location of \( 45 \, \mathrm{mm} \). At \textcolor{green!50!black}{(10,0)}, the maximum discrepancy increases to \( 15\% \), also at \( 45 \, \mathrm{mm} \). For \textcolor{violet}{(10,10)}, the maximum discrepancy decreases to \( 9\% \) at \( 40 \, \mathrm{mm} \). Finally, at the \textcolor{blue}{edge (15,0)}, the largest discrepancy of \( 20\% \) is observed at \( 45 \, \mathrm{mm} \). These locations and values correspond to the curve colors shown in Figure~\ref{fig_ratio_plot_data_nominal_model}.

\section{Optimization and Results}
Fitting optimizes the magnetization parameters of the neodymium pieces in the COMSOL model to minimize discrepancies with measured data. The objective is to determine the 147 parameters that yield the best agreement. Once optimized, this model enables interpolation and extrapolation of the magnetic field map.

\subsection{Objective Function}
\label{subsec:objective_function}
The fitting process minimizes a chi-squared (\( \chi^2 \)) objective function, defined as:
\[
\chi^2 = \sum_{i=1}^{N_{\text{DataPoints}}} \left( \frac{(b_{x,i} - b_{x,\text{pred},i})^2}{\sigma^2} + \frac{(b_{y,i} - b_{y,\text{pred},i})^2}{\sigma^2} + \frac{(b_{z,i} - b_{z,\text{pred},i})^2}{\sigma^2} \right),
\]
where \( b_{x,i}, b_{y,i}, b_{z,i} \) are measured field components, \( b_{x,\text{pred},i}, b_{y,\text{pred},i}, b_{z,\text{pred},i} \) are COMSOL predictions, \( \sigma = 0.01 \, \mathrm{T} \) and $\sum_{i=1}^{N_{\text{DataPoints}}}$ = 4095, i.e., the total number of data points in the AP-STD measured map.\\

The fitting process starts with inputting an initial parameters file in COMSOL and running simulations to calculate magnetic field maps. The workflow iteratively updates parameters via the ``SCAN" algorithm from Minuit2 until the \(\chi^2\) is minimized. The final fit parameters are used to generate a high-resolution 1 mm map with this fitted COMSOL model for further analysis. COMSOL ensures Maxwell's equations are satisfied in the fit.

\begin{figure}[h!]
    \centering
    \includegraphics[width=0.8\linewidth]{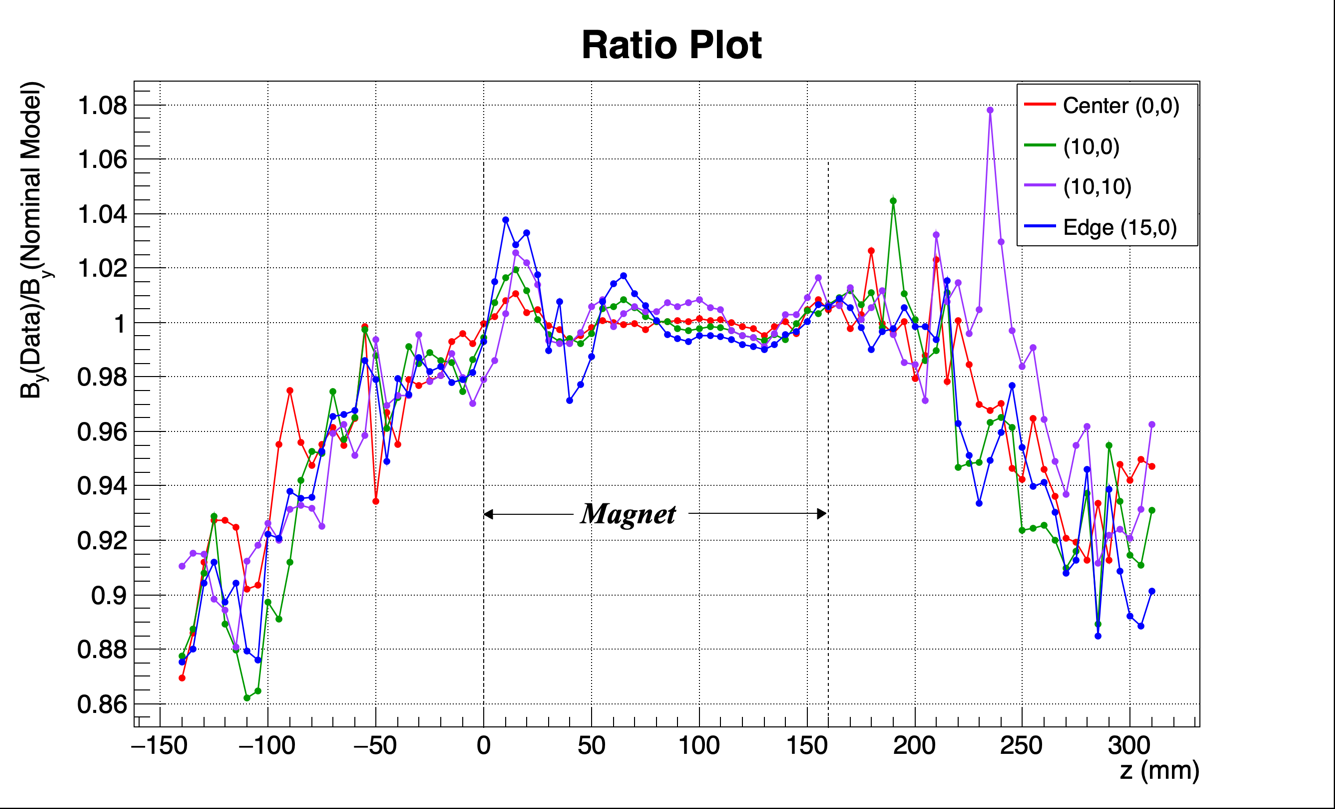}
    \caption{Ratio of data to the \textbf{fitted} COMSOL model at key positions. Compare with Figure~\ref{fig_ratio_plot_data_nominal_model}.}
    \label{fig_ratio_data_fit}
\end{figure}

\begin{wrapfigure}{r}{0.3\linewidth}
    \centering
    \vspace{-30pt}
    \includegraphics[width=\linewidth]{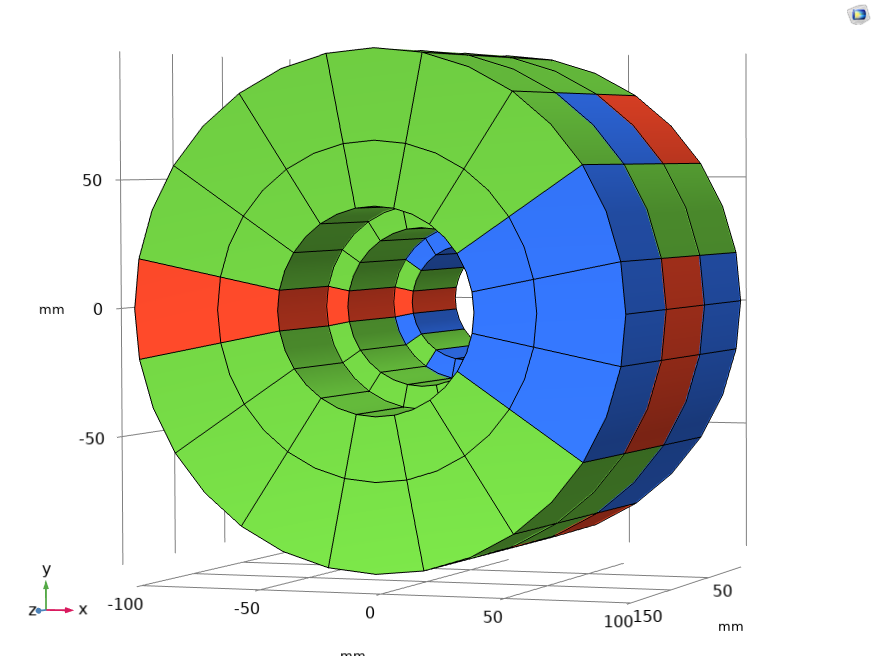}
    \caption{Relative percentage changes across magnet segments.}
    \label{fig:relative_change_diagram}
\end{wrapfigure}

\textbf{Results:}  
In Figure~\ref{fig_ratio_data_fit} we can see that the optimization significantly reduces discrepancies. At the edges, the maximum discrepancy decreases from \(\sim 20\%\) to \(\sim 5\%\), while along the central axis, it reduces from \(\sim 11\%\) to \(\sim 1\%\). Figure~\ref{fig:relative_change_diagram} highlights relative changes in the resultant magnetizations after the fit, relative to the nominal magnetizations in each of the 48 magnet segments. Green (\(< 10\%\)), blue (10–20\%), and red (\(> 20\%\)) denote varying deviations.\\

The final optimized parameters allow the generation of a 1 mm magnetic field map for EMPHATIC analysis, ensuring high spatial resolution and accuracy, while also extending the map beyond the measured volume.

\printbibliography

\end{document}